\documentclass[%
 reprint,
superscriptaddress,
 amsmath,amssymb,
pra,
]{revtex4-2}
\usepackage{url}
\usepackage{braket}
\usepackage{graphicx}
\usepackage{dcolumn}
\usepackage{bm}
\usepackage{color,soul}
\usepackage{hyperref}
\usepackage{newfloat}
\DeclareFloatingEnvironment[name={FIG.}]{appfigure}

\hypersetup{
    colorlinks=true,
    linkcolor=blue,
    filecolor=magenta,      
    urlcolor=cyan,
}
\begin{document}

\preprint{APS/123-QED}

\title{Gate-based spin readout of hole quantum dots with site-dependent $g-$factors}

\author{Angus Russell}%
\thanks{these authors contributed equally}%
\affiliation{
Department of Physics, SUPA, University of Strathclyde, Glasgow G4 0NG, United Kingdom}
\author{Alexander Zotov}%
\thanks{these authors contributed equally}%
\affiliation{
Department of Physics, SUPA, University of Strathclyde, Glasgow G4 0NG, United Kingdom}
\author{Ruichen Zhao}%
\affiliation{
School of Electrical Engineering and Telecommunications, University of New South Wales, Sydney, New South Wales 2052, Australia}
\author{Andrew S. Dzurak}%
\affiliation{
School of Electrical Engineering and Telecommunications, University of New South Wales, Sydney, New South Wales 2052, Australia}
\author{M. Fernando Gonzalez-Zalba}%
\affiliation{
Quantum Motion Technologies, Nexus, Discovery Way, Leeds, LS2 3AA, United Kingdom
}
\author{Alessandro Rossi}
\email{alessandro.rossi@strath.ac.uk}
\affiliation{
Department of Physics, SUPA, University of Strathclyde, Glasgow G4 0NG, United Kingdom}
\affiliation{National Physical Laboratory, Hampton Road, Teddington TW11 0LW, United Kingdom
}

\begin{abstract}
The rapid progress of hole spin qubits in group IV semiconductors has been driven by their potential for scalability. This is owed to the compatibility with industrial manufacturing standards, as well as the ease of operation and addressability via all-electric drives. However, owing to a strong spin-orbit interaction, these systems present variability and anisotropy in key qubit control parameters such as the Land\'e $g-$factor, requiring careful characterisation for reliable qubit operation. Here, we experimentally investigate a hole double quantum dot in silicon by carrying out spin readout with gate-based reflectometry. We show that characteristic features in the reflected phase signal arising from magneto-spectroscopy convey information on site-dependent $g-$factors in the two dots. Using analytical modeling, we extract the physical parameters of our system and, through numerical calculations, we extend the results to point out the prospect of conveniently extracting information about the local $g-$factors from reflectometry measurements.
\end{abstract}

\maketitle


\section{\label{sec:Intro}Introduction}
Spin quantum bits (qubits) based on hole quantum dots realized in silicon and germanium are promising systems for the realization of large scale quantum computers~\cite{floris,scappucci21,anasua21}. This is because they can combine two key ingredients. Firstly, they can be manufactured via well-established industrial processes~\cite{mfgz21,intel22}. Secondly, qubit operation can be achieved through all-electric control pulses~\cite{maurand16,watzinger,nico20}, which avoids the need for cumbersome hardware overhead dedicated to producing a.c. magnetic fields (antennas) or local field gradients (micromagnets).\\\indent
A distinctive feature of hole qubits is a sizable spin-orbit interaction (SOI), which ultimately  enables the mentioned electric-field-driven spin control. The SOI significantly affects singlet-triplet ($S-T$) qubits because, depending on its origin, may lead to $S-T_0$ mixing, $S-T_-$ mixing or both~\cite{Li15,jirovec22,Crippa2019}. The former scenario takes place when the mixing of heavy-hole and light-hole bands occurs, which in turn makes the Land\'e $g-$factor susceptible to local fluctuations of confinement, strain and material chemistry~\cite{jirovec21,liles18,liles21}, which in practice makes it site-dependent. The latter situation arises when the spin-orbit length is comparable to the quantum dot lateral size leading to spin-flip tunnel coupling~\cite{golovach}. This may result in spin-blockade lifting and state leakages to the detriment of $S-T$ qubit performance. The traditional way of characterizing the details of SOI at play in qubits often requires transport measurements and the ability to resolve small spin-blockade leakage currents~\cite{Li15,Mutter_2020}. In the perspective of scaling up the qubit count, characterization techniques, such as gate-based radio-frequency (rf) reflectometry~\cite{vigneau}, which do not rely on transport, would be preferable because they are compatible with the readout of individual qubit cells in large arrays~\cite{schaal,ruffino,mfgz21}.\\\indent
Here, we use gate-based rf readout to experimentally investigate the $S-T$ manifold in a hole-based silicon double quantum dot (DQD) via magneto-spectroscopy. We establish that the associated phase signal conveys information which can be used to extract the values of site-dependent local $g-$factors in the two dots. Furthermore, we show that the expected rf readout signal originating from spin-flip tunneling events is incompatible with our observations, and conclude that these events do not play a significant role in our experimental conditions. Finally, we generalise our methodology with numerical calculations. These allow us to elucidate the interplay among tunnel coupling, thermal energy and local $g-$factors on the characteristics of the phase response.
\section{\label{sec:Methods}Methods}
\begin{figure}[t]
\includegraphics[scale=0.54]{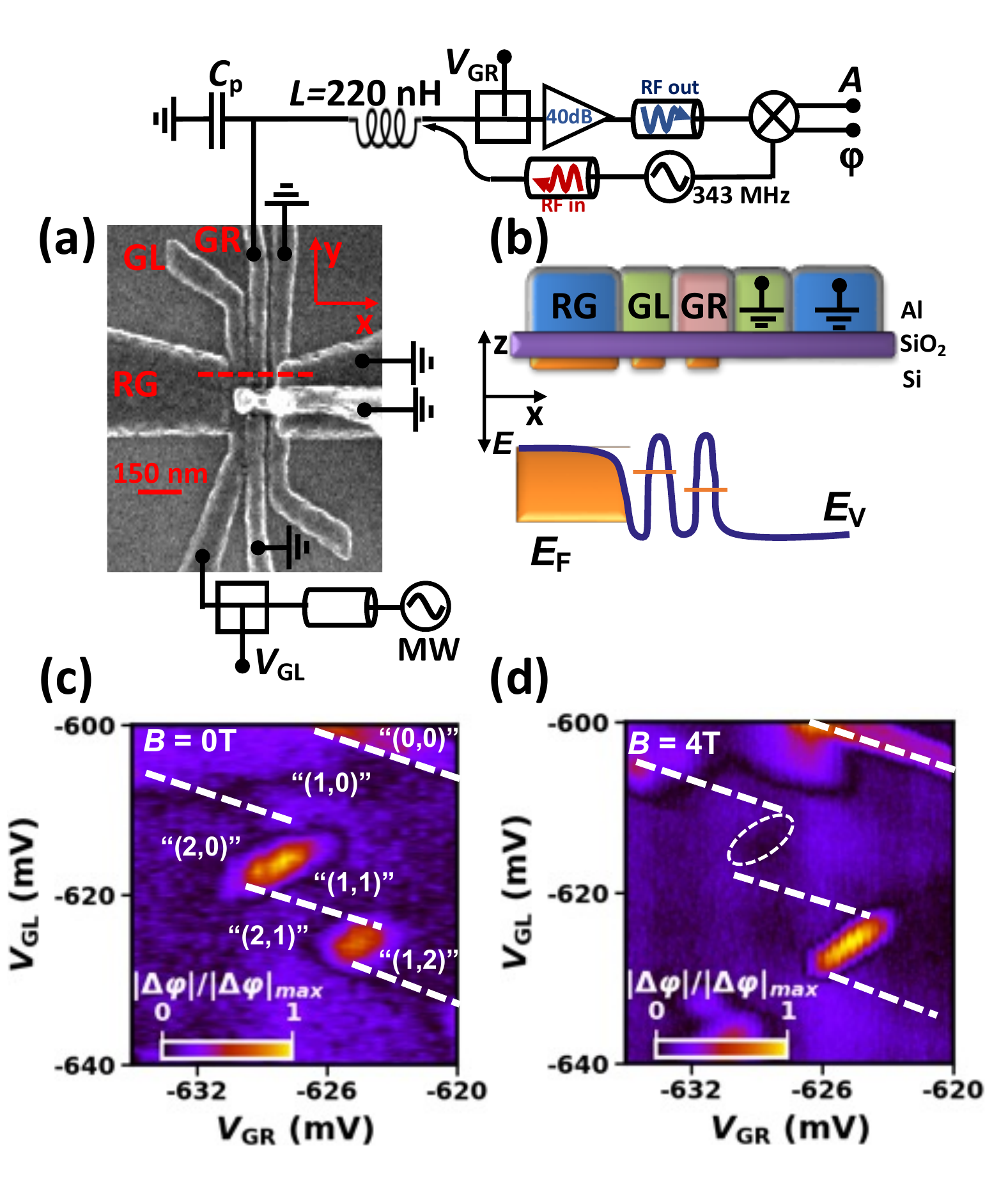}
\caption{(a) SEM image of a device similar to the one used in the experiments and schematic view of the measurement set-up. Gates that are not labeled have been kept at ground potential. GR and GL are connected to bias tees to superimpose d.c. voltages to rf signals for dispersive readout and microwave (MW) excitation, respectively. (b) Cross-sectional view of the device stack (top) and potential profile (bottom) along the dashed line of panel (a). Regions in orange indicate hole layer accumulation. Gate colors indicate different metallization layers. (c) Normalized phase response as functions of dc voltages applied to GL and GR at $B=0$~T and $V_\textup{RG}=-1.5$~V. Dashed lines are guides to the eye indicating dot-reservoir charge transitions. Relative DQD hole occupancy is reported as per the discussion in the text. (d) Phase response for the same voltages as in (c) at $B=4$~T. Dashed oval highlights the region where an even parity ICT vanished.}
\label{fig:1}
\end{figure}

The sample used is a p-type metal-oxide-semiconductor (MOS) field-effect transistor fabricated on a near-intrinsic natural silicon substrate (resistivity $>10$~k$\Omega\cdot$cm). Three layers of Al/Al$_y$O$_x$ gates are patterned with electron-beam lithography and deposited on an 8-nm-thick SiO$_2$ gate oxide~\cite{angus,jove, rossi17}. A scanning electron micrograph (SEM) image of the metal gate stack of a device similar to the one used in the experiments is shown in Fig.~\ref{fig:1}(a). Upon application of negative dc voltages to individual gate electrodes, one can locally accumulate a layer of holes or form tunnel barriers at the Si/SiO$_2$ interface. By cooling down the device in a dilution refrigerator with a base temperature of $30$~mK, we form a DQD as illustrated in Fig.~\ref{fig:1}(b) and control its occupancy via dc voltages applied to gates GL and GR. To readout the polarisation state of the DQD~\cite{vigneau}, we connect GR to a lumped-element LC resonator formed by a surface mount inductor ($L=220$~nH) and the parasitic gate-to-ground capacitance ($C_\textup{p}$), resulting in a resonant frequency $f_\textup{r}~\approx~343$ MHz. The measurement set-up used for rf reflectometry is schematically represented in Fig.~\ref{fig:1}(a), where the base-band phase and amplitude of the signal reflected by the resonator are labeled $\varphi$ and $A$, respectively. Hereafter, we shall focus on the resonator's phase response.

\begin{figure}[t]
\includegraphics[scale=0.43]{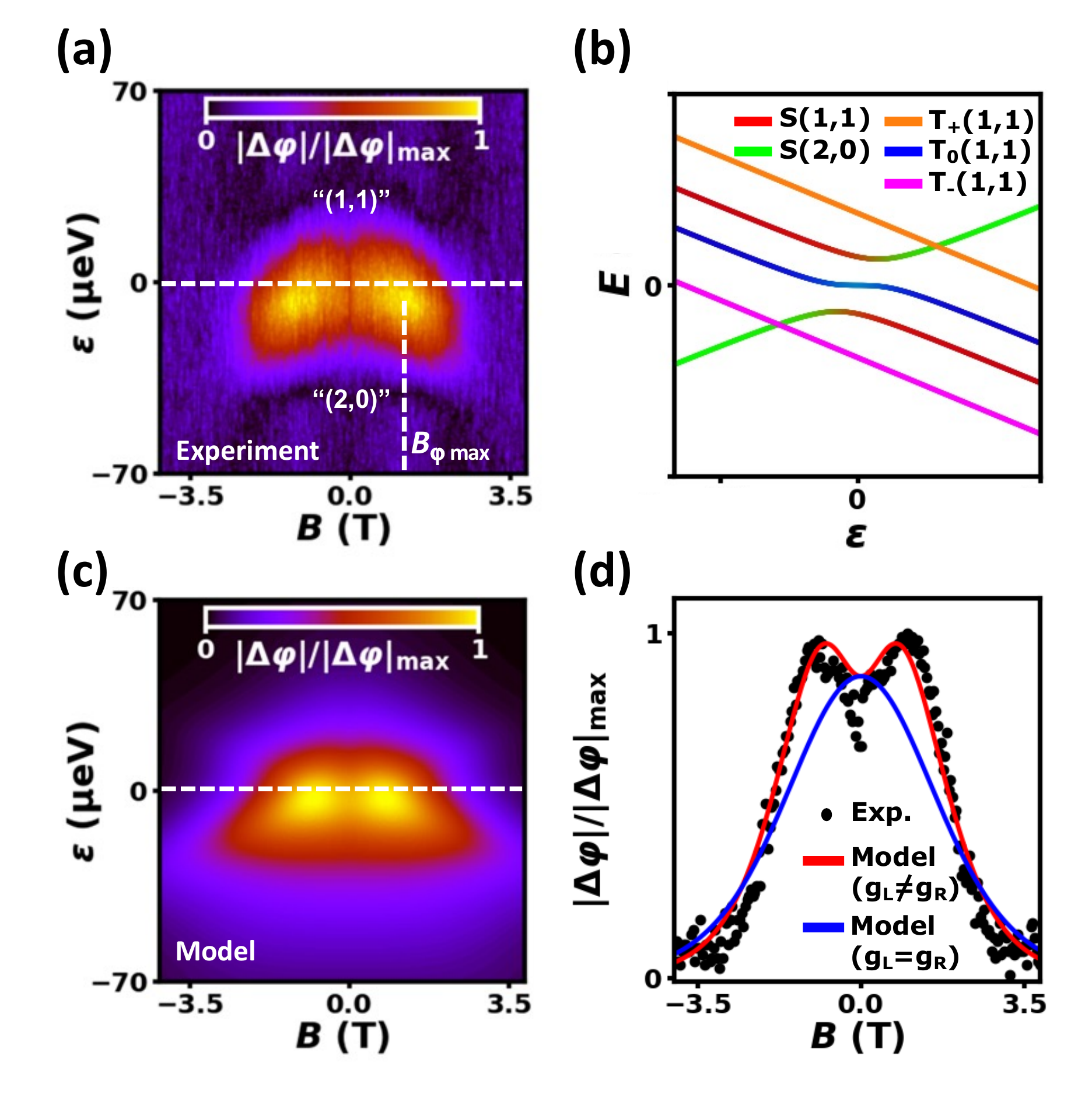}
\caption{(a) Normalised phase response measured as functions of detuning energy and magnetic field. Horizontal dashed line indicates the zero detuning line cut. Vertical dashed line highlights the value of magnetic field, $B_{\varphi max}$, at which the maximum of phase signal occurs. (b) DQD energy levels at an even parity ICT with site-dependent $g-$factor values and a finite $B-$field, as derived from the Hamiltonian in the text. (c) Calculated normalised phase response as functions of detuning energy and magnetic field. Parameters used are: $g_{\rm L}=0.49$, $g_{\rm R}=0.28$, $t=10.6~\mu$eV, $T_{\rm DQD}=250~$mK. (d) Measured phase response (black dots) as a function of $B-$field at zero detuning, as obtained from the line cut shown in panel (a). Red solid trace is a fit to the experimental data yielding the $g-$factor values used in panel (c). For comparison, the monotonic blue trace is calculated by imposing equal $g-$factor values: $g=g_{\rm R}+g_{\rm L}=0.77$.}
\label{fig:2}
\end{figure}
\section{\label{sec:Results}Results}
Figure~\ref{fig:1}(c) shows a charge stability map as a function of $V_\textup{GL}$ and $V_\textup{GR}$ \cite{wiel}. The measured phase shifts highlight the boundaries where stable charge configurations occur. We label the hole occupancy in the two dots ``($N_\textup{L}$, $N_\textup{R}$)", where $N_\textup{L}$ ($N_\textup{R}$) is the number of valence holes in the left (right) quantum dot. The quotation marks highlight the fact that, due to disorder in the device \cite{geer, hyst, thorbeck, supp}, we are unable to define an absolute number of carriers in each dot but are interested in identifying the parity of charge transitions. In order to discriminate between even and odd transitions, we apply a magnetic field ($B$) in the plane of the hole layer, i.e. the $xy$-plane of panel (a), to determine which inter-dot charge transition (ICT) phase response is affected by Pauli spin blockade~\cite{schroer}. As shown in Fig.~\ref{fig:1}(d), one transition completely disappears at $B=4$~T revealing its even occupancy status, whereas a transition with one additional hole persists as expected for odd parity occupancy. These observations led to the charge attributions shown in panel (c).\\\indent
The measured dependence of the ``($1$,$1$)"$-$``($2$,$0$)" transition on $B-$field and detuning, $\varepsilon$, is displayed in Fig.~\ref{fig:2}(a). Note that raw data are taken as a function of $V_\textup{GL}$ and converted to energy using the interdot gate lever-arm, $\alpha$ (see Appendix~\ref{sec:alpha-eps}). At high values of magnetic field, the phase signal decreases in intensity, eventually vanishing whilst shifting towards more negative values of detuning. These effects stem from the fact that for increasing magnetic field the lower triplet state, $T_-$, decreases in energy and eventually becomes the ground state. The shift in the position of the peak as a function of $B$ tracks the crossing point between the $T_-$ and the lower singlet state, $S$, see Fig.~\ref{fig:2}(b). Interestingly, the phase response evolves non-monotonically for increasing $B$, reaching a point of maximum signal strength at $B_\textup{$\varphi$max} =1.28$~T at finite detuning, as reported in Fig.~\ref{fig:2}(a). In previous experimental investigations on electronic DQD systems, the $B$-field dependence of the resonator response was reported to decrease monotonically with increasing $B$~\cite{schroer,betz15}.  We ascribe the non-monotonic effect to the SOI in holes, particularly different $g$-factor values between the left and right quantum dots~\cite{Crippa2018,Crippa2019,Mutter_2020}, as discussed next. 
\begin{figure}[b]
\includegraphics[scale=0.95]{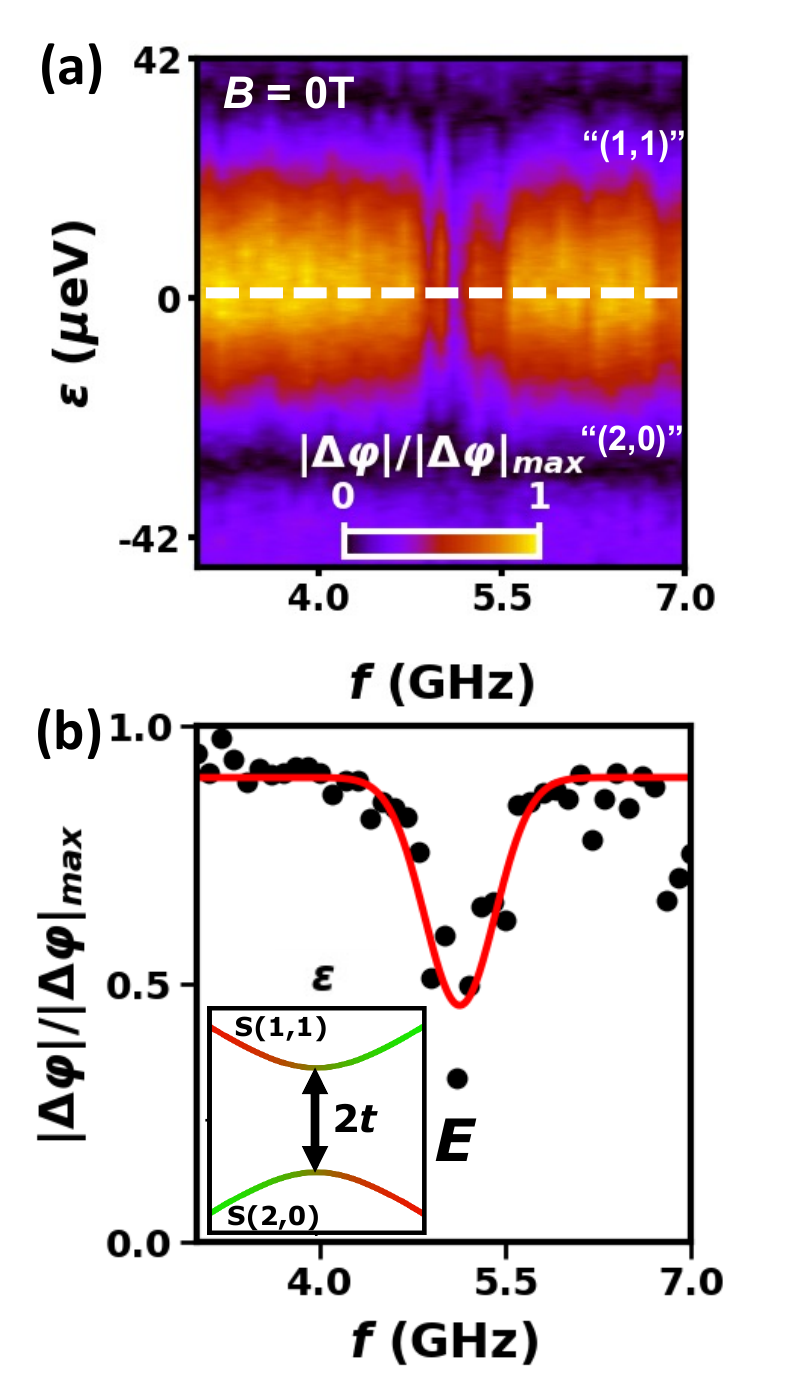}
\caption{(a) Normalized resonator response measured in the vicinity of the even parity ICT as functions of detuning and MW frequency for $B=0$~T and $V_\textup{RG}=-1.5$~V. The dashed line is a guide to the eye to highlight the zero detuning data cut of panel (b). (b) Phase response as a function of MW frequency at $\varepsilon=0$ (dots) and Gaussian fit (red line). Inset: Energy levels of the DQD near $\varepsilon=0$ for the even parity ICT in the absence of a magnetic field (degenerate triplet states are omitted for visual clarity).}
\label{fig:3}
\end{figure}
\\\indent In order to explain the observed magnetic field dependence, we build an analytical model of the resonator response based on a Hamiltonian in the singlet-triplet basis $\{T_+(1,1),T_0(1,1),T_-(1,1),S(1,1),S(2,0)\}$:

\begin{widetext}
\begin{equation}
\label{eq:hamil_glgr}
H=\left( \begin{array}{ccccc}   
-\frac{1}{2}\varepsilon+ \frac{1}{2}(g_\textup{L}+g_\textup{R})\mu_\textup{B} B & 0 & 0 & 0 & 0\\  
0 & -\frac{1}{2}\varepsilon & 0&\frac{1}{2}(g_\textup{L}-g_\textup{R})\mu_\textup{B} B &0\\  
0 & 0 & -\frac{1}{2}\varepsilon- \frac{1}{2}(g_\textup{L}+g_\textup{R})\mu_\textup{B} B &0 &0\\  
0 & \frac{1}{2}(g_\textup{L}-g_\textup{R})\mu_\textup{B} B & 0&-\frac{1}{2}\varepsilon  &t\\  
0 & 0 & 0&t &\frac{1}{2}\varepsilon \\  
\end{array}\right)
\end{equation}
\end{widetext}
where $t$ is the interdot tunnel coupling, $\mu_\textup{B}$ is the Bohr magneton and $g_\textup{L}$ ($g_\textup{R}$) is the Land\'e $g-$factor in the left (right) dot for the given orientation of the magnetic field. Figure~\ref{fig:2}(b) shows a representative energy spectrum as a function of detuning for finite $B$ and $g_\textup{L} \neq g_\textup{R}$. Besides the usual singlet anticrossing due to tunnel coupling, one can notice that another anticrossing arises between the singlet states and $T_\mathrm{0}$ due to electric dipole-induced coupling~\cite{Crippa2019}, given that the Zeeman splitting in the two dots is not the same. We argue that the coupling between the $T_0(1,1)$ and the ground singlet state produces a contribution to the dispersive readout signal which is compatible with the observed non-monotonic $B-$field dependence. Note that in the chosen basis set, we deliberately neglect spin-orbit coupling terms, $t_\textup{SO}$, that would lead to spin-flip tunneling events, such as transitions $T_-(1,1)-S(2,0)$ or $T_+(1,1)-S(2,0)$. We maintain that, due to Landau Zener (LZ) excitation~\cite{anasuaPRB}, these terms would not produce a measurable dispersive signal in experimental conditions compatible with ours. As discussed in Appendix~\ref{sec:LZ}, this is the case as long as $t_\textup{SO}<0.70~\mu$eV. Furthermore, we model the effect of a larger $t_\textup{SO}$ in the absence of LZ excitation and conclude that it would produce a signal with strikingly different magnetic field dependence with respect to our experimental observations, see Appendix~\ref{sec:tso}. These calculations reveal that, unlike shown in Fig.~\ref{fig:2}(a), the phase signal would not vanish for large Zeeman splittings. This is due to the fact that the $T_-$ state acquires a curvature that generates an additional signal, tracking the $S-T_-$ anticrossing as the $B-$field increases~\cite{han,theo21}. \\\indent
In order to model the reflectometry readout signal, we use a semiclassical approach based on quantum capacitance in the adiabatic limit~\cite{Crippa2019,betz15}. In particular, we calculate the contribution to the overall phase response, $\Delta\varphi$, from the five states, $E_\textup{i}$, each weighted assuming that they are populated according to a Boltzmann distribution with an effective temperature $T_\textup{DQD}=250$~mK  (this estimate is based on characterisation of similar devices carried out with the same experimental set-up).  The component of the phase signal attributable to each state, $\Delta\varphi_i$, is quantified in terms of a quantum capacitance contribution to the resonator circuit and reads $\Delta\varphi_i\propto C_{Q_i}=-(e\alpha)^2(\partial^2E_\textup{i}/\partial\varepsilon^2)$, where $e$ is the elementary charge~\cite{betz15}. The measured phase signal is an average over several cycles of the rf probe signal at the resonant frequency during which each state is partially thermally populated~\cite{schroer}. Hence, the overall readout signal can be written as $\Delta\varphi=\sum_i\langle\Delta\varphi_i\rangle\propto \sum_i C_{Q_i}\frac{e^{-E_\textup{i}/k_\textup{B}T_\textup{DQD}}}{\sum_i e^{-E_\textup{i}/k_\textup{B}T_\textup{DQD}}}$, where $k_\textup{B}$ is the Boltzmann constant.
\\\indent
We use this model to calculate the phase response in an analytical form and to extract the $g-$factor values at each dot site via a fit to the experimental data. In order to have $g_\textup{L}$ and $g_\textup{R}$ as the only free parameters of the model, we extract $t$ from the MW spectroscopy data shown in Fig.~\ref{fig:3}(a).
In particular, we observe that by driving the system with continuous MW excitation in the absence of a magnetic field, the phase signal at the even parity ICT is significantly reduced at a given frequency, $f_0$. This indicates that a MW-driven transition between the ground and excited branch of the singlet states is occurring~\cite{urdampilleta,ezzo21}. In fact, the state promotion results in a signal reduction because, due to antisymmetric curvatures, each branch contributes with opposite sign to the overall phase signal. The maximum signal reduction takes place when the MW photon energy matches $2t$, see Fig.~\ref{fig:3}(b). By means of a Gaussian fit, we extract $f_0=5.12$~GHz and consequently $t=\frac{hf_0}{2}=10.6~\mu$eV, where $h$ is the Planck's constant. We then feed this into our model and fit the magnetospectroscopy data of Fig.~\ref{fig:2}(a) to the analytical solution for $\varepsilon=0$, see Fig.~\ref{fig:2}(d). The extracted values for the g-factors are $g_\textup{L}=0.49$ and $g_\textup{R}=0.28$. These are compatible with values measured in similar MOS devices for which it has been reported that the $g-$factor strongly depends on dot occupancy, gate voltage and magnetic field orientation~\cite{liles18,liles21}. Note that our model makes no particular distinction between the two quantum dots and, therefore, the extracted values can also be swapped between left and right side.  With the knowledge acquired for all experimental parameters, we can calculate the full magnetic field response, as depicted in Fig.~\ref{fig:2}(c). It is evident that the model reproduces the main experimental features including the vanishing of the readout signal at high field due to spin blockade, and, crucially, the non-monotonic evolution of the phase response for increasing values of $B$, with the phase response reaching a maximum at a critical finite field, $B_{\varphi\textup{max}}$. We can now understand the origin of this effect. As observed in systems with uniform $g-$factors~\cite{schroer,betz15}, for large values of $B$, state $T_-(1,1)$ becomes the ground state with dominant occupation probability. The linearity of this state's energy dependence provides a negligible contribution to the quantum capacitance of the system and explains the decreasing and eventually vanishing resonator response. However, in the case of site-dependent $g-$factors, for relatively small Zeeman splittings, the additional bending acquired by the singlet-ground state due to the coupling with $T_0(1,1)$, contributes significantly to the overall quantum capacitance. Depending on the values of $B$ and $\varepsilon$, this contribution may either add up or subtract from the overall phase response (see also Appendix~\ref{sec:non-mon}). The net result is that $|\Delta\varphi|$ increases to a local maximum before vanishing for increasing $B$. The value $B_{\varphi\textup{max}}$ arises from the interplay of the different energy scales that affect the signal contributions from each state, namely $t$, $k_\textup{B}T_\textup{DQD}$ and $\frac{1}{2}(g_\textup{L}\pm g_\textup{R})\mu_\textup{B}B$. \\\indent
\begin{figure*}[]
\includegraphics[scale=0.85]{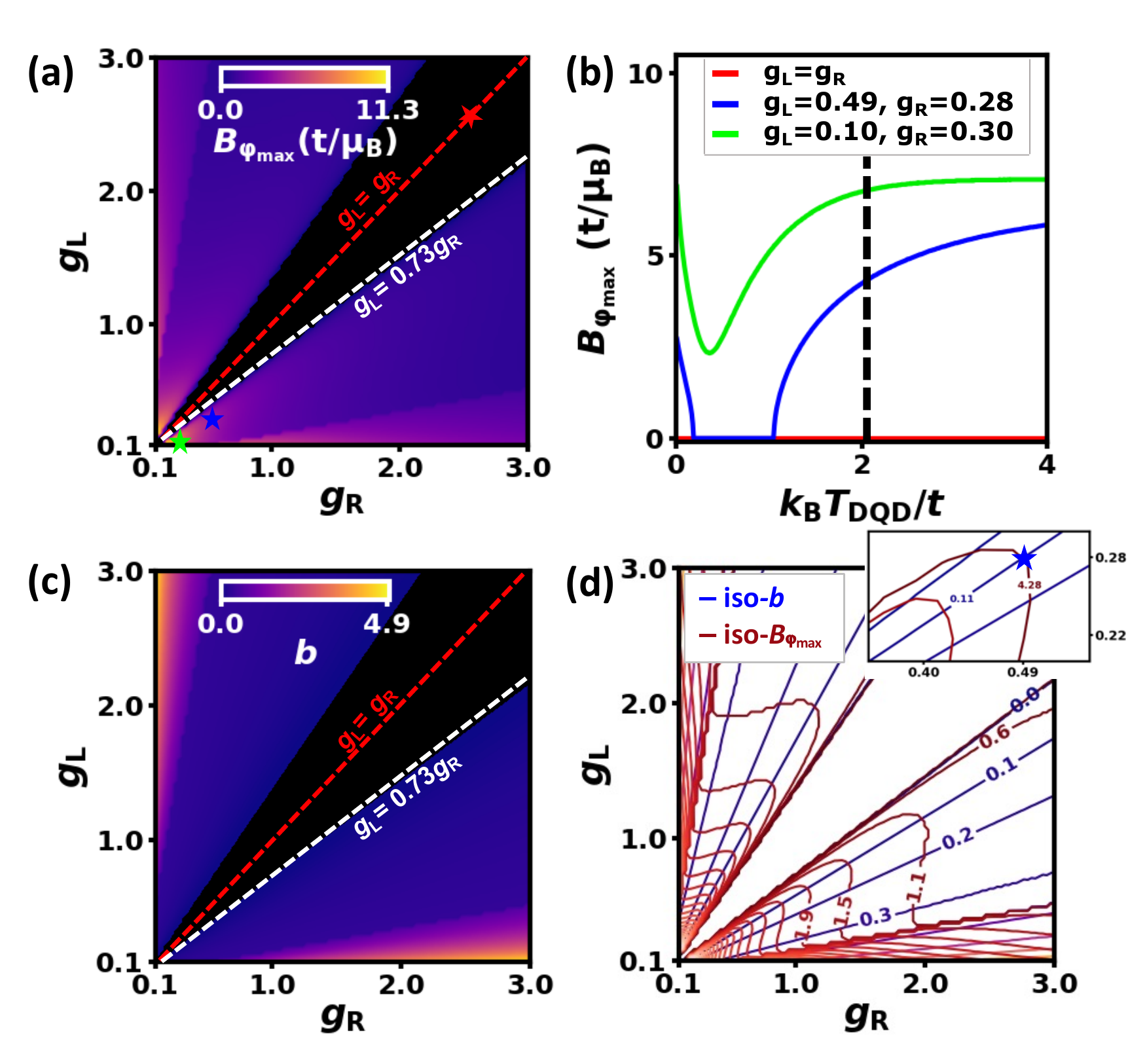}
\caption{(a) Numerical calculations of $B_{\varphi max}$ as functions of $g_{ \rm L}$ and $g_{ \rm R}$ for $k_{\rm B} T_{\rm DQD}/t=2.03$. Dashed lines are guides to the eye to indicate the loci of $g_{ \rm L}=g_{ \rm R}$ (red) and $g_{ \rm L}=0.73g_{ \rm R}$ (white). Stars highlight selected $g-$factor values studied in panel (b) and are colour-coded accordingly. The blue star indicates the $g-$factors extracted from the experiments. (b) Calculated $B_{\varphi max}$ as a function of the ratio of thermal and tunnel coupling energies for different combinations of $g-$factor values. The experimental conditions are met at the intersection between the blue trace and the vertical dashed line. (c) Numerical calculations of $b$ as as functions of $g_{ \rm L}$ and $g_{ \rm R}$ for $k_{\rm B} T_{\rm DQD}/t=2.03$. (d) Traces of iso$-B_{\varphi max}$ (red) and iso$-b$ (blue) as functions of $g_{ \rm L}$ and $g_{ \rm R}$ obtained from the calculations of panels (a) and (c), respectively. Inset: Enlarged view highlighting the point of intersection (star) for the $g-$factors extracted by fitting to the analytical model of Fig.~\ref{fig:2}(c).}
\label{fig:4}
\end{figure*}
In order to reinforce that the non-monotonicity is the signature of site-dependent $g-$factors and elucidate the implications of varying energy scales, we have developed a numerical model that calculates $B_{\varphi\textup{max}}$ as a function of the main experimental parameters. In fact, $B_{\varphi\textup{max}}$ is a proxy for the type of $B-$field dependence because $B_{\varphi\textup{max}}=0$ represents a monotonic dependence whereas $B_{\varphi\textup{max}}\neq 0$ indicates a non-monotonic one. Figure~\ref{fig:4}(a) shows the dependence of $B_{\varphi\textup{max}}$ on the $g-$factor values. The calculations reveal that $B_{\varphi\textup{max}}=0$ for $0.73\leq \frac{g_{\rm L}}{g_{\rm R}}\leq 1$ which ultimately represents the locus of a monotonic decrease of readout signal as a function of increasing magnetic field. Note that the boundaries of this condition depend on the ratio $k_{\rm B} T_{\rm DQD}/t$, and we have deliberately chosen $k_{\rm B} T_{\rm DQD}/t=2.03$ to mimic our experimental situation (see Appendix~\ref{sec:kT/t} for the effect of varying this ratio). It is evident that site-dependent $g-$factors are a necessary condition for the non-monotonicity to arise albeit not always sufficient. We also note that the value of $B_{\varphi\textup{max}}$ rapidly increases as the $g-$factors approach small values like those indicated by a green and a blue star in Fig.~\ref{fig:4}(a). This makes the occurrence of the non-monotonicity significantly easier to notice in the experiments like in our case (blue star) for which $B_{\varphi\textup{max}}=1.28~$T. By contrast, for larger $g-$factors, although these may satisfy the condition for non-monotonic behaviour, the value of $B_{\varphi\textup{max}}$ may be on the order of few tens of mT and may become harder to detect. As mentioned, the ratio between thermal and tunnel coupling energies plays an important role. In Fig.~\ref{fig:4}(b), we show its effect on $B_{\varphi\textup{max}}$ for three representative choices of $g-$factors. The calculations further confirm that for DQDs with uniform $g-$factors the $B-$field dependence is always monotonic irrespective of temperature and tunnel coupling, as $B_{\varphi\textup{max}}=0$. Interestingly, for the values of $g-$factors found in our experiments (blue trace), $B_{\varphi\textup{max}}$ can be either finite or zero depending on the value of the energy ratio. Note also that for particular values of $g-$factors a non-monotonic dependence can arise irrespective of $k_{\rm B} T_{\rm DQD}/t$, as the absence of zeros in the green trace demonstrates. Another useful parameter to take into account is the relative enhancement in phase signal achieved at $B=B_{\varphi\textup{max}}$. We call this the ``boost factor", $b$, and define it as $b=\frac{\Delta\varphi_\textup{max}-\Delta\varphi_0}{\Delta\varphi_0}$, where $\Delta\varphi_\textup{max}=\Delta\varphi(B=B_{\varphi\textup{max}})$ and $\Delta\varphi_0=\Delta\varphi(B=0)$ at the same detuning point. In Fig.~\ref{fig:4}(c), we calculate the dependence of $b$ on $g-$factor values. The plot shows a consistency with the boundaries of monotonic/non-monotonic response of panel (a), as expected by the fact that the condition $b=0$ is representative of monotonic $B-$field dependence. The calculations also reveal that the signal boost is more significant the larger the difference in $g-$factor values between dots is. Interestingly, one could use the calculations in panels (a) and (c) in tandem to extract the pair of $g-$factor values at play in the DQD. As we show in Fig.~\ref{fig:4}(d), the intersection points between contour lines of iso$-B_{\varphi\textup{max}}$ and iso$-b$ provide such values as coordinates directly from a visual inspection of the diagram. In general, one could use diagrams of the kind of panel (d) as chart maps whose contours are calculated based on the knowledge of the DQD temperature and tunnel coupling. The specific chart for the experimental setting is then used to identify the local $g-$factors from the intersection point between relevant iso-traces, provided prior knowledge of $B_{\varphi\textup{max}}$ and $b$ from reflectometry readout. As shown in the inset of Fig.~\ref{fig:4}(d), the $g-$factor values extracted in this way are consistent with those obtained by fitting to the experimental data in Fig.~\ref{fig:2}(d), confirming a consistency between numerical and analytical calculations. Note that this approach for extracting site-dependent $g-$factors is alternative to the established routes that require spin blockade lifting via electric dipole spin resonance~\cite{Crippa2018} or the measurement of the frequency of coherent spin oscillations~\cite{jirovec22}. With respect to these established techniques, the method described here may be advantageous in simplifying the experimental set-up requirements because it does not rely on sophisticated hardware to generate the pulse sequences needed for coherent qubit control.
\section{\label{sec:Disc}Conclusion}
We have performed magneto-spectroscopy of a hole-based Si DQD through reflectometry readout. We have demonstrated  that a non-monotonic dependence of the  phase signal with respect to magnetic field is the signature of site-dependent $g-$factors. We stress that the large value observed for $B_{\rm \varphi max}=1.28$~T is incompatible with a trivial magnetic field dependence originating from a superconducting-normal transition in the Al bond wires that form part of the resonant circuit \cite{Caplan1965, krollMag,petersson2012circuit}. Such trivial non-monotonicity would have instead produced a local peak at the transition critical field value of a few tens mT, as already reported in Ref.~\cite{schroer} (see also the Supplemental Material \cite{supp}).  By contrast, through analytical and numerical calculations, we have linked the details of the non-monotonic signal response to a general set of properties of the DQD system and extracted its $g-$factors. Such understanding based on gate-based readout could be useful to engineer the optimal conditions to operate spin qubits in devices for which transport measurements are not possible, such as 1D or 2D DQD arrays. In an ideal scenario, one would want to work with the maximum readout signal to enhance signal-to-noise ratio and facilitate single-shot readout protocols~\cite{west,Pakkiam,zheng}. However, this has to be compatible with relatively small $B-$field operation to preserve good coherence and relaxation times. To this end, our calculations have shown that the tuning of the $g-$factors could be effective in widely altering both $b$ and $B_{\rm \varphi max}$. Such tuning could be achieved with several strategies, e.g. by controlling the direction of the magnetic field, the gate-induced electric field, the DQD charge occupancy or the interface strain\cite{Crippa2018,liles18,liles21}.\\   
\begin{acknowledgments}
The authors thank O. Henrich, S.P. Giblin and M. Kataoka for useful discussions. A. Rossi and M.F. Gonzalez-Zalba acknowledge support from the UKRI Future Leaders Fellowship Scheme (Grant agreements: MR/T041110/1, MR/V023284/1). 
\end{acknowledgments}
\begin{appendix}

\section{\label{sec:alpha-eps}Energy Detuning and gate lever arm}
\begin{figure}[h]
		\includegraphics[scale=0.59]{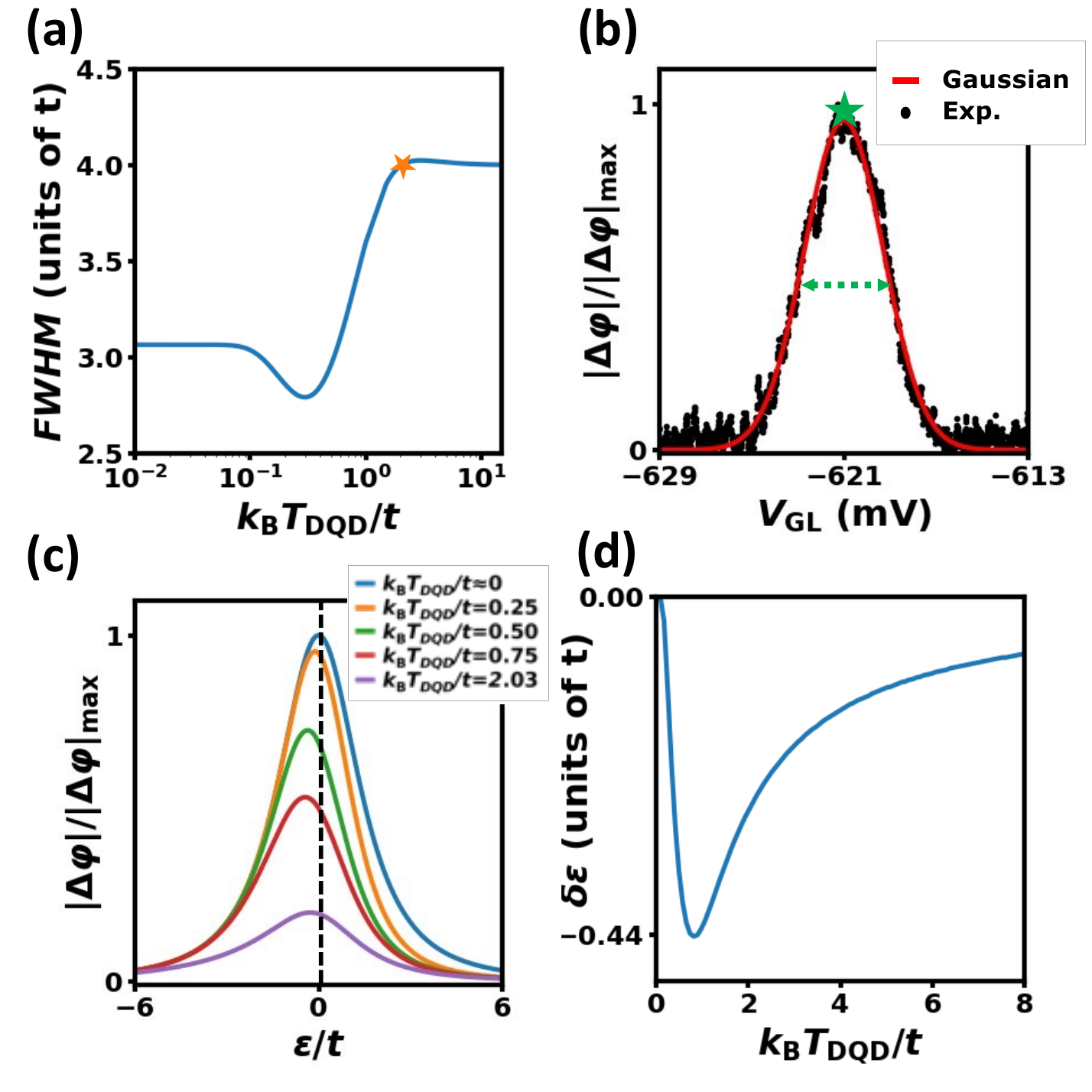}
		\caption{(a) Calculated FWHM as a function of thermal energy normalised to tunnel coupling for the phase signal of an even parity ICT as modeled in the main text. A star indicates the experimental condition. (b) Normalised phase response for the even parity ICT studied in the main text, as a function of gate voltage at $B=0$~T. Experimental data (dots) are fitted to a Gaussian function (solid line) from which the position of the peak maximum (star) and FWHM (dashed arrows) are extracted. (c) Calculated normalised phase response as a function of detuning for different temperatures and zero magnetic field (all energies are normalised to tunnel coupling). Dashed vertical line highlights the zero detuning axis, which coincides with a maximum of the phase signal only at zero temperature. (d) Calculated energy shift away from zero detuning as a function of temperature at the maximum of the phase signal, as per traces in panel (c).}
		\label{appfig:lever}
	\end{figure}

 To convert from voltage into energy detuning we use $\varepsilon  = \alpha(V_{\rm GL} - V_0)$, where $\alpha$ is the inter-dot gate lever-arm, $V_{\rm GL}$ is the voltage applied to GL and $V_0$ is the voltage that corresponds to zero detuning. We determine $\alpha$ without a direct measurement by utilising the relationship between the full width at half maximum (FWHM) of the even parity ICT phase response at $B=0$~T and the effective device temperature, $T_{\rm DQD}$, as discussed in Ref. \cite{Crippa2019}. As reported in the main text, we assume $T_{\rm DQD}= 250~\rm mK$ and extract $t=10.6~\rm{\mu eV} $ from MW spectroscopy experiments. Hence, one can obtain the FWHM in volts by fitting a Gaussian to a cut along the detuning axis of the magneto-spectroscopy data at $B=0$~T shown in Fig. \ref{appfig:lever}(b), which can then be compared to the FWHM obtained from the model in units of $t$ in Fig.~\ref{appfig:lever}(a). The factor relating these two quantities is the interdot lever-arm,
 \begin{equation}
	\label{eq:lever}
		\alpha= \frac{\rm FWHM_{\rm model}(t)}{\rm FWHM_{\rm data}(\rm mV)}=10.49~\mu\rm{eV}/\rm{mV}
	\end{equation}
We now turn to evaluate $V_0$. Figure \ref{appfig:lever}(c) shows the phase signal calculated at $B=0$~T for different values of $k_{\rm B} T_{\rm DQD}/ t$. It can be seen that the maximum of the phase response does not occur at $\varepsilon=0$ for finite temperatures. This shift can be attributed to the asymmetry of the degenerate triplet states' energy around zero detuning. At positive detuning, the triplet states have lower energy than at negative detuning and are therefore populated to a greater extent for a finite value of $k_{\rm B} T_{\rm DQD}/ t$. This increased state population reduces the overall phase signal only in the positive detuning sub-space leading to a shift of the maximum phase response towards negative detuning. This effect is also evident in Fig.~\ref{appfig:a3}. The shift, $\delta\varepsilon$, is plotted as a function of $k_{\rm B} T_{\rm DQD}/ t$ in Fig. \ref{appfig:lever}(d). We convert the expected shift for our system parameters into  voltage and find $V_0$ as
	\begin{equation}
		V_0 = V_{\rm max} + \frac{\delta \varepsilon}{\alpha}= -620.72~\rm{mV}
	\end{equation}
	where $V_{\rm max} = -621\rm{mV}$ is obtained from the Gaussian fit in Fig. \ref{appfig:lever}(b).
 \begin{figure}[t]
\includegraphics[scale=0.75]{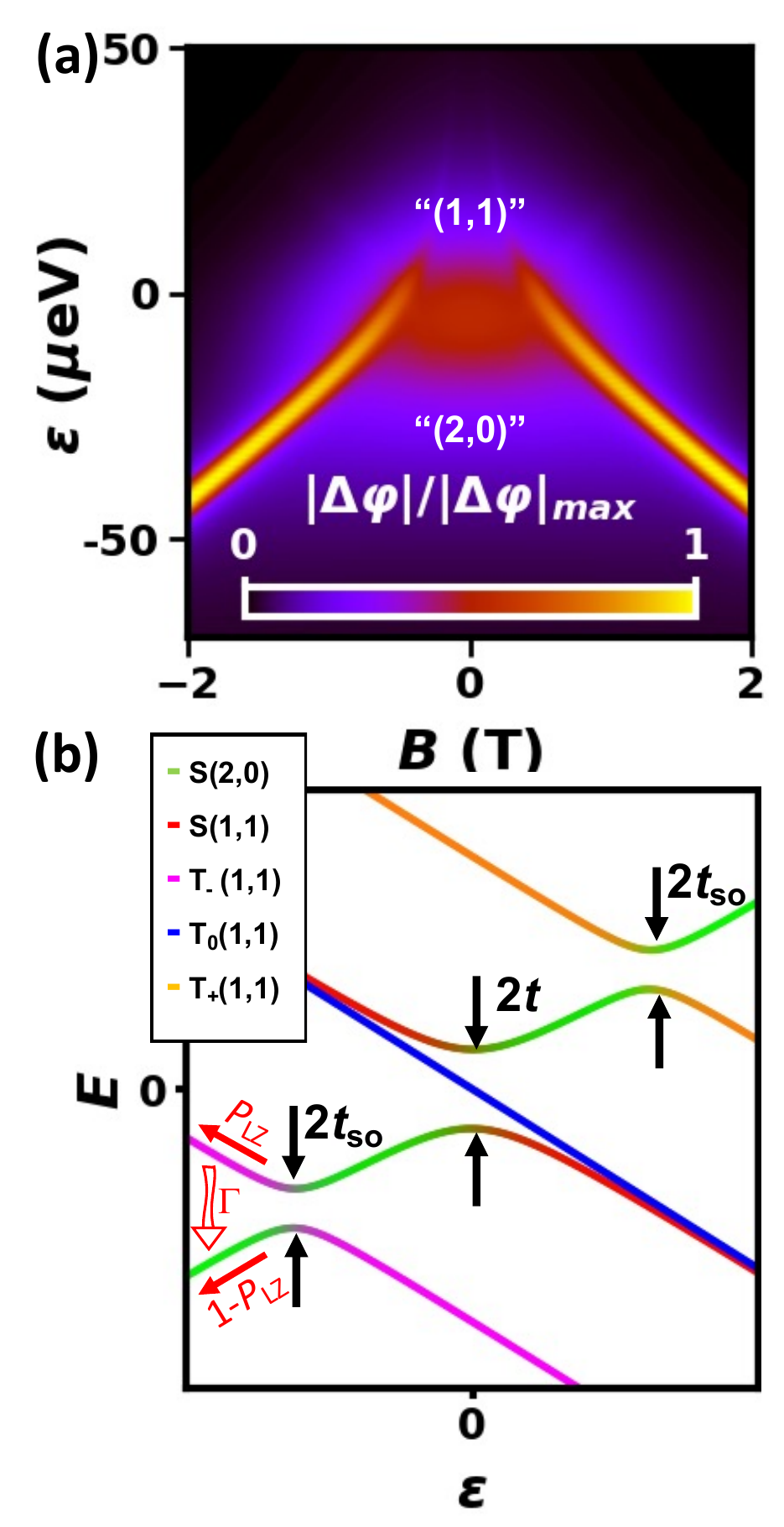}
\caption{(a) Calculated normalised phase response as functions of detuning energy and magnetic field. The values of parameters used in the model are $g=\frac{g_L+g_R}{2}=0.39$, $t=10.6~\mu$eV, $t_\textup{SO}=1.0~\mu$eV, $T_\textup{DQD}=250$~mK. (b) DQD energy levels at an even parity ICT in the vicinity of zero detuning and finite value of $B$. For pictorial clarity, energy scales are exaggerated with respect to the values of panel (a).}
\label{appfig:tso}
\end{figure}
\section{\label{sec:LZ}Effect of Landau Zener excitation}
Let us consider the anticrossing between $ST_-$ caused by spin-flip tunnelling terms, $t_\textup{SO}$, and establish the conditions for which the rf drive at frequency $f_\textup{r}=343$~MHz would produce Landau Zener (LZ) excitation. This situation is of interest because, for a large transition probability ($P_\textup{LZ}$) stemming from sufficiently long $ST_-$ relaxation rate ($\Gamma=1/T^{ST_-}_1$) and coherence time ($T^{ST_-}_2$), there may be a vanishingly small quantum capacitance contribution to the resonator response from this anticrossing, see Fig.~\ref{appfig:tso}(b).
In fact, the signal-to-noise ratio ($SNR_\textup{LZ}$) of the resonator response would be reduced because of a reduced occupation probability of the ground state and an increased probability of occupation of the excited state, which subtracts from the ground state contribution because of its anti-symmetric curvature. This  results in $SNR_\textup{LZ}=SNR(1-2P_\textup{LZ})$, where $SNR$ is assumed to be equal to the value at the tunnel coupling anticrossing for simplicity. In order to calculate $P_\textup{LZ}$, we use the QuTiP toolbox~\cite{qutip} and evaluate the stationary probability distribution in the multipassage regime. The condition for which the $ST_-$ anticrossing would not produce a measurable phase signal is $SNR_\textup{LZ}<1$, which ultimately leads to an upper bound for the value of the spin-flip term
\begin{equation}
\label{eq:upper}
		t_\textup{SO}<0.70~\mu \textup{eV}
\end{equation}
For this calculation, we have extracted $SNR=14.9$ from the data in Fig.~\ref{appfig:lever}(b), $T_2\approx 700$~ns from Ref~\cite{jirovec21}. and the amplitude of the drive at the gate $A_\textup{rf}=4.4~\mu$eV from the knowledge of rf amplitude and the lever arm of eq.~\ref{eq:lever}.
\\\indent The condition to be satisfied for the relaxation rate is 
\begin{equation}
\label{eq:lower}
		T^{ST_-}_1>\frac{1}{2f_r}= 1.46~\textup{ns} 
	\end{equation}
Typical relaxation rates of relevant transitions for hole DQDs are in the region of several $\mu$s~\cite{Crippa2018,Crippa2019}, which suggests that in our system the LZ-induced excitation will not relax to the ground state within several periods of the drive.\\\indent  
In conclusion, both conditions set by inequalities~\ref{eq:upper}, \ref{eq:lower} are likely to be met in our experiments and we consider, therefore, justified the assumption to neglect $t_\textup{SO}$ terms in the model Hamiltonian of Eq.~\ref{eq:hamil_glgr}.

\section{\label{sec:tso}Dispersive response with sizable spin-flip tunnelling terms}
In order to further corroborate our decision to neglect $t_\textup{SO}$ terms in Eq.~\ref{eq:hamil_glgr}, we present the results of calculations obtained with sizable $t_\textup{SO}$. Furthermore, we assume $g=\frac{g_\textup{L}+g_\textup{R}}{2}=0.39$ for both left and right dot. This is because we want to explicitly check whether spin-flip tunnelling alone could provide an alternative explanation to site-dependent Land\'e $g-$factors for our experimental observations. We use a Hamiltonian in the same single-triplet basis as in the main text, which reads,
\begin{widetext}
\begin{equation}
\label{eq:hamil_tso}
H'=\left( \begin{array}{ccccc}   
-\frac{1}{2}\varepsilon+g\mu_\textup{B} B & 0 & 0 & 0 & t_\textup{so}\\  
0 & -\frac{1}{2}\varepsilon & 0&0 &0\\  
0 & 0 & -\frac{1}{2}\varepsilon-g\mu_\textup{B} B &0 &t_\textup{so}\\  
0 & 0 & 0&-\frac{1}{2}\varepsilon  &t\\  
t_\textup{so} & 0 & t_\textup{so}&t &\frac{1}{2}\varepsilon \\  
\end{array}\right)
\end{equation}
\end{widetext}
The results of our numerical calculations for varying detuning, $\varepsilon$, and magnetic field, $B$, are shown in Fig.~\ref{appfig:tso}(a). For illustrative purposes, we have used $t_\textup{SO}=1~\mu$eV, i.e. larger than the upper bound calculated in ineq.~\ref{eq:upper}. 
The resonator phase response, $\Delta\varphi$, is calculated by quantum capacitance contributions of each eigenstate weighted by its occupation probability according to a Boltzmann distribution, as explained in the main text. There is a striking discrepancy with the experimental data of Fig.~\ref{fig:2}(a). In particular, the phase response does not vanish at high values of $B$. This can be understood from the fact that, for increasing $B$, $T_-(1,1)$ eventually becomes the ground state and it keeps contributing significantly to the readout signal thanks to the bending acquired via the coupling with $S(2,0)$, see Fig.~\ref{appfig:tso}(b). This is ultimately a mechanism through which spin blockade is lifted due to spin-flip tunnelling~\cite{Li15}. In principle, for negative detuning and large magnetic field a contribution to the overall quantum capacitance should be measurable in the event of sizable $t_\textup{SO}$. Given that we have not observed this feature in the relevant data-sets of Fig.~\ref{fig:1}(d) and Fig.~\ref{fig:2}(a), we have neglected this term's contribution in our model. However, we do not exclude that spin-flip tunneling terms may play a more significant role for different orientations of the magnetic field~\cite{han}.
\begin{figure*}[]
\includegraphics[scale=1.0]{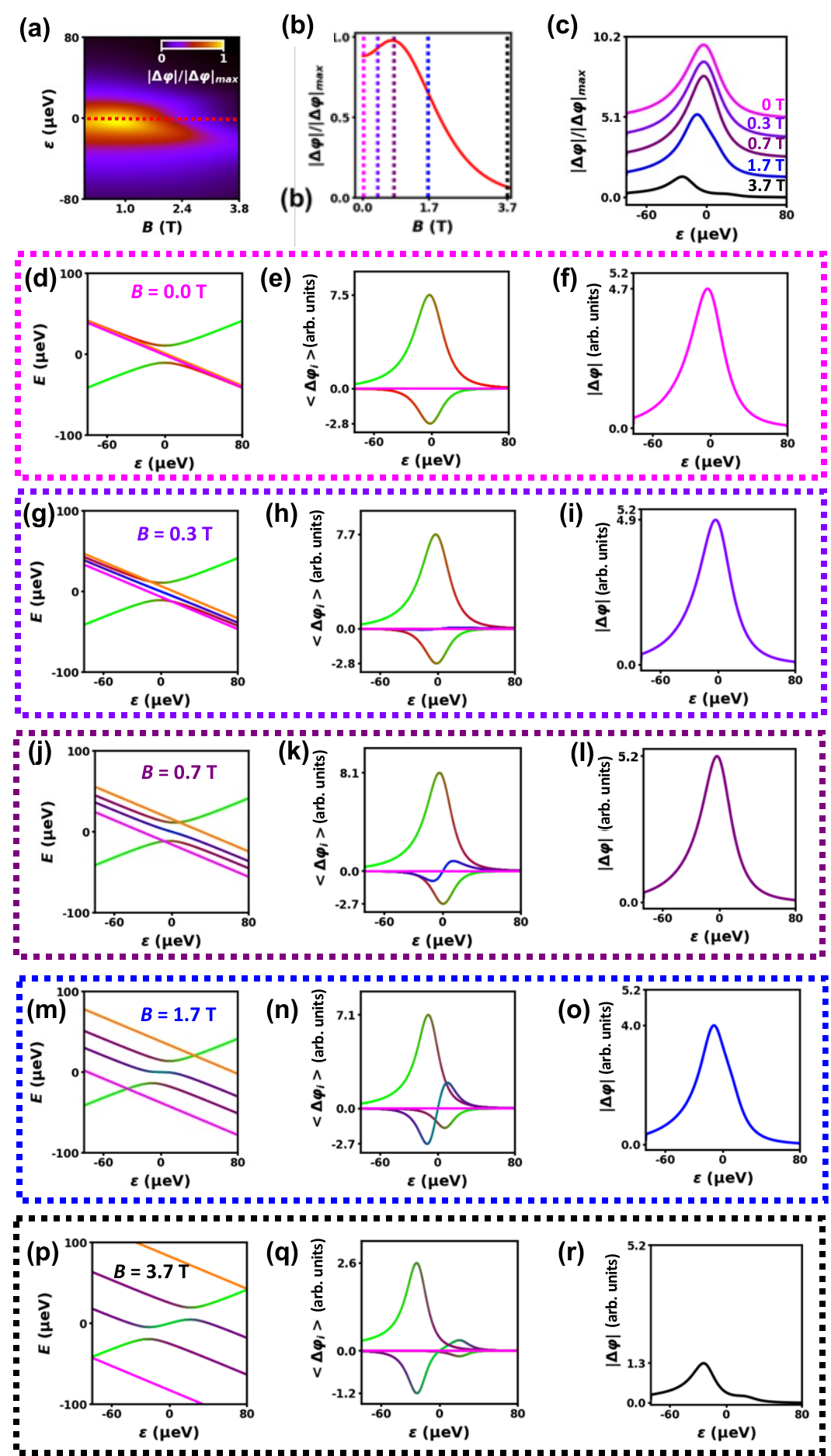}
\caption{(a) Calculated phase response for an even parity ICT as functions of $B$ and $\varepsilon$. Parameters used are $g_\textup{L}=0.493$, $g_\textup{R}=0.279$ , $t=10.6~\mu$eV, $T_\textup{DQD}=250$~mK. Dotted line shows the position of zero detuning used for panel (b). (b) Phase response at zero detuning as a function of $B$. Dotted lines show the $B-$field values selected for the calculations of the individual and overall phase responses. (c) Normalised phase response as a function of detuning for the $B$ values indicated by dashed lines in panel (b). Traces are offset vertically for clarity. (d), (g), (j), (m), (p) Energy eigenvalues of five lowest energy levels as a function of detuning at $B=0.0~$T, $B=0.3~$T, $B=0.7~$T, $B=1.7~$T, $B=3.7~$T, respectively. The color coding is the same as in Fig.~\ref{appfig:tso}. (e), (h), (k), (n), (q) Individual eigenstate phase responses weighted to the Boltzmann occupation probability for the selected magnetic field values. The color coding uniquely associates the signal contribution to the relevant eigenstate. (f), (i), (l), (o), (r) Overall phase responses for each of the selected magnetic field values.}
\label{appfig:a3}
\end{figure*}
\begin{figure}[]
\includegraphics[scale=0.65]{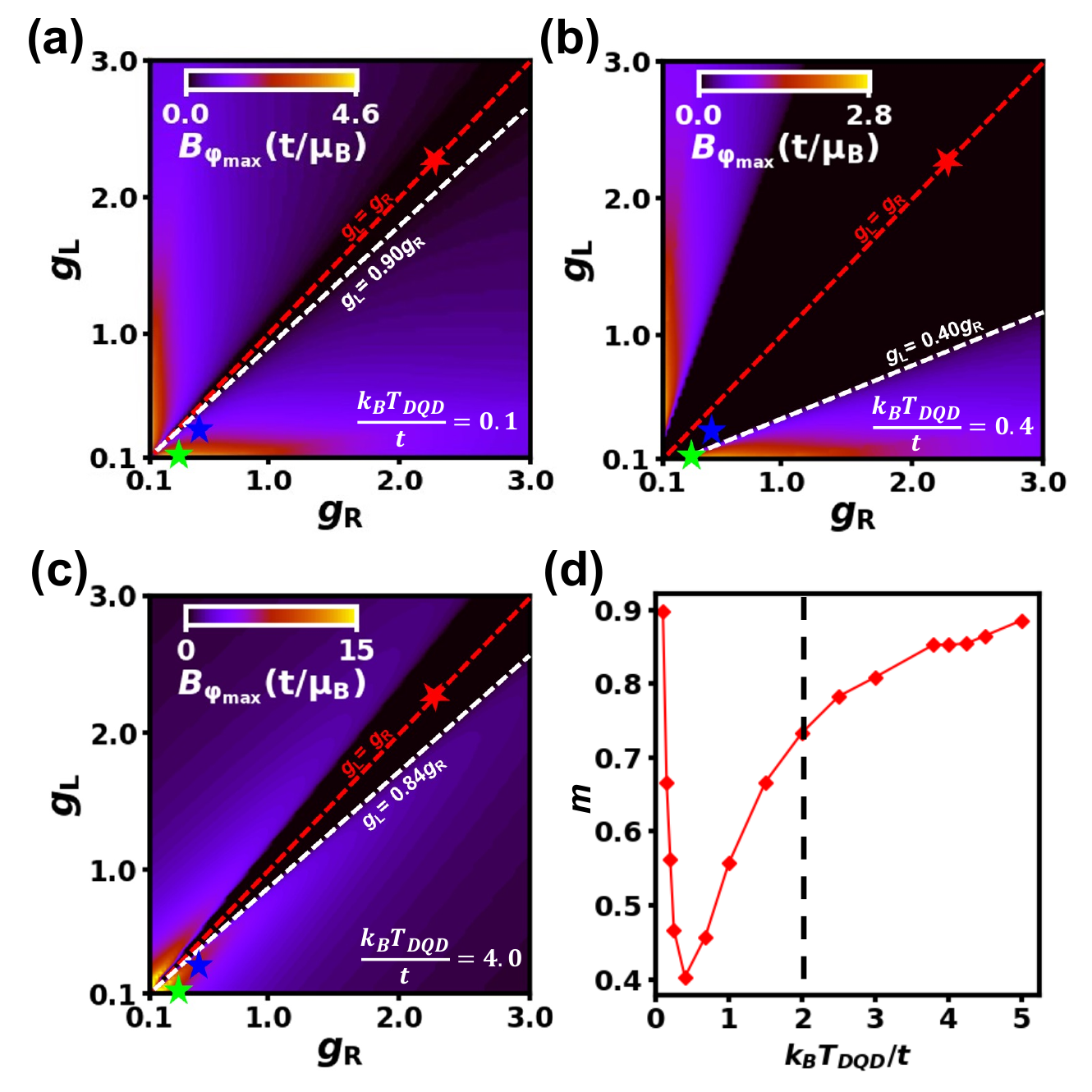}
\caption{Numerical calculations of $B_{\varphi\textup{max}}$ as functions of $g_\textup{L}$ and $g_\textup{R}$ for (a) $k_\textup{B}T_\textup{DQD}/t=0.1$, (b) $k_\textup{B}T_\textup{DQD}/t=0.4$, (c) $k_\textup{B}T_\textup{DQD}/t=4.0$. Dashed lines are guides to the eye to indicate the loci of $g_\textup{L}=g_\textup{R}$ (red) and $g_\textup{L}=m g_\textup{R}$ (white). Blue star highlights the $g-$factor values extracted from fits  to  the experiment. They could result in either monotonic or non-monotonic phase response depending on $k_\textup{B}T_\textup{DQD}/t$. Green star highlights a combination of $g-$factor values for which the response is non-monotonic irrespective of the value of $k_\textup{B}T_\textup{DQD}/t$. (d) $m$ as a function of $k_\textup{B}T_\textup{DQD}/t$ extracted from fits to the $B_{\varphi\textup{max}}=0$  locus boundary. The dashed line indicates the experimental condition discussed in the main text.}
\label{appfig:a4}
\end{figure}
\section{\label{sec:non-mon}Non-monotonic phase response}
In order to illustrate the origin of the non-monotonic $B-$field dependence for the phase response, we have calculated the individual contributions, $<\Delta\varphi_{\rm i}>$, from the energy levels weighted for the appropriate occupation probability. We have also calculated the relevant overall phase signal, $|\Delta\varphi|$, for representative values of $B$.  Figure~\ref{appfig:a3}(a) shows the overall phase response as per Fig.~\ref{fig:2}(c) but limited to positive magnetic field. By considering a slice of this data-set at zero detuning, the non monotonic dependence can be seen, as depicted in Fig.~\ref{appfig:a3}(b) where vertical dotted lines are used to indicate the magnetic field values chosen to illustrate the initial increase in signal amplitude followed by a fall-off.\\\indent
At $B=0$~T, the signal contribution from the ground singlet state dominates albeit sligthly reduced by a negative contribution from the excited singlet (see Fig.~\ref{appfig:a3}(d-f)). No contributions from the degenerate triplet states is to be expected due to lack of curvature. As the field increases to $B=0.3~\rm T$, the contribution from the ground singlet increases because its curvature is enhanced by the anti-crossing with the $T_0$ state (see Fig.~\ref{appfig:a3}(g-i)), which ultimately occurs because of the site-dependent nature of the $g-$factors. As a result, $|\Delta\varphi|$ increases and reaches a maximum at approximately $B=0.7~\rm T$ in our calculations (see Fig.~\ref{appfig:a3}(j-l)). We stress that, in the absence of such an anticrossing caused by $g_{\rm L} \neq g_{\rm R}$, the phase contribution from the curvature of the singlet would not increase with $B$ and, therefore, a non-monotonicity would not arise. For larger values of field, the phase response begins a roll-off for two reasons. Firstly, the contribution from $T_0$ subtracts from the contribution of the singlet for negative detuning (see Fig.~\ref{appfig:a3} (m-o)) where it had shifted to for reasons already discussed in Appendix~\ref{sec:alpha-eps}. Secondly, the occupation of the ground singlet is reduced with respect to the occupation of $T_-$, which eventually turns into the most favourable energy level (see Fig.~\ref{appfig:a3} (p-r)). At even higher values of $B$, this latter effect explains the complete disappearance of the phase signal because of the lack of curvature in the $T_-$ detuning dependence.

\section{\label{sec:kT/t} Effect of $\frac{k_\textup{B}T_\textup{DQD}}{t}$ on signal monotonicity}
In the main text we have shown that, in order to attain a non-monotonic phase response,  the ratio between the $g-$factor values has to satisfy the inequality $g_\textup{L}/g_\textup{R}<0.73$. However, this condition strongly  depends on experimental conditions, such as the temperature of the system, $T_\textup{DQD}$, and the interdot tunnel coupling, $t$. In Fig.~\ref{appfig:a4}(a-c), we have calculated the value of $B_{\varphi\textup{max}}$ as functions of $g_\textup{L}$ and $g_\textup{R}$ for three representative values of the ratio $k_\textup{B}T_\textup{DQD}/t$. By fitting the boundary of the parameter space defined by $B_{\varphi\textup{max}}=0$ to a straight line with slope $m$, one can write the condition defining the non-monotonic behavior in the following parametrized form: $g_\textup{L}/g_\textup{R}<m$, with $m$ dependent on $k_\textup{B}T_\textup{DQD}/t$. In  Fig.~\ref{appfig:a4}(d), we plot $m$ as extracted for different values of $k_\textup{B}T_\textup{DQD}/t$. The  value of $m$ reaches a  minimum at $k_\textup{B}T_\textup{DQD}/t=0.4$ corresponding to $T_\textup{DQD}=50$~mK in the case of $t=10.6~\mu$eV, which we extracted from MW spectroscopy experiments shown in Fig.~\ref{fig:3} . As shown in Fig.~\ref{appfig:a4}(b), this is a situation where the g-factors parameter space leading to non-monotonic phase response is the narrowest. Another interesting aspect is that for vanishingly small temperature, $m$ increases rapidly leading to more relaxed requirements on the values of the $g-$factors to achieve non-monotonic behavior, see Fig.~\ref{appfig:a4}(a). The fact that this effect can also arise at zero temperature, i.e. when the occupation probability of the ground state is near-unity, is a consequence of an initial increase in curvature of the ground singlet due to the anticrossing with $T_0$, resulting in an enhanced quantum capacitance contribution. 
\end{appendix}

%

\pagebreak
\widetext
\newpage
\begin{center}
\textbf{\large SUPPLEMENTARY INFORMATION\\[2in]Gate-based spin readout of hole quantum dots with site-dependent $g-$factors}
\end{center}
\setcounter{equation}{0}
\setcounter{figure}{0}
\setcounter{table}{0}
\setcounter{section}{0}
\setcounter{page}{1}
\makeatletter
\renewcommand{\theequation}{S\arabic{equation}}
\renewcommand{\thefigure}{S\arabic{figure}}
\renewcommand{\thesection}{S\Roman{section}}



\newpage
\thispagestyle{plain}

\section{\label{sec:multi-dot}Multiple Double Quantum Dot system}
\begin{figure}[h]
\includegraphics[scale=0.45]{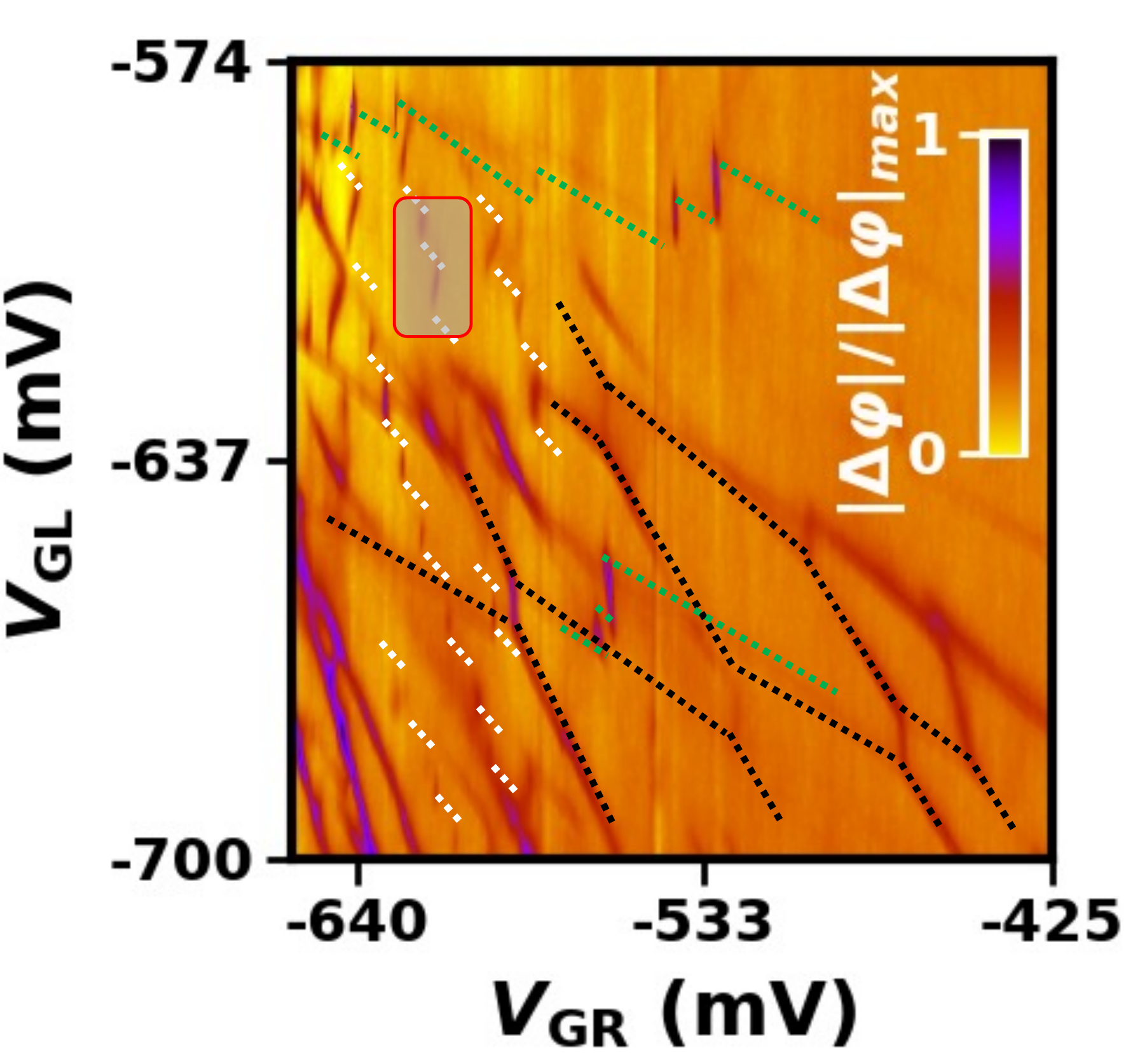}
\caption{Resonator phase response as a function of dc voltages applied to GL and GR at $B=0$~T and $V_\textup{RG}=-1.55$~V. All other gates are grounded. Dashed lines are guides to the eye indicating dot-reservoir charge transitions color coded for three different DQDs simultaneously operating in the device. The grey shaded area identifies the gate voltage space investigated in the main text.}
\label{sfig:largemap}
\end{figure}

Figure~\ref{sfig:largemap} shows a charge stability map acquired by scanning $V_\textup{GL}$ and $V_\textup{GR}$ within relatively large ranges. The data contain numerous features that can be attributed to either dot-reservoir charge transitions or inter-dot charge transitions (ICT).  It is possible to assign ICTs to different double quantum dots (DQDs) by looking at their intensity in the phase response, given that this relates to the relevant gate lever arm characteristic of each DQD. Another indicator is the slope of both the ICTs and the reservoir transitions which relates to the specific capacitive couplings of each DQD. Hence, we can identify at least three separate DQD systems coupled to the single reservoir formed under gate RG, which all undergo charge transitions for the scanned gate voltages. In particular, we highlighted with green and white dotted lines the left-dot-to-reservoir charge transitions of two distinct DQDs. These lines undergo vertical shifts whenever an ICT takes place. A third DQD system is identified by reservoir transitions indicated with black lines, which are arranged in the familiar honeycomb pattern~\cite{wiel}. In this case, we infer that the inter-dot tunnel barrier is sufficiently transparent to allow co-tunnelling events~\cite{geer} between the right dot and the reservoir, resulting in both dots being able to exchange holes with the reservoir~\cite{hyst}.\\
\indent The origin of multiple DQDs in our device can be ascribed to several factors. For example, in the absence of a screening gate, a hole layer can form under the entire area of the readout gate (i.e. near, as well as  far away from the reservoir) because the device is operated in accumulation-mode regime~\cite{rossi17}. This may lead to the formation of disordered DQDs arising from puddles of holes formed by unintended variation of gate width or strain effects due to thermal coefficient mismatch~\cite{thorbeck}.\\
\indent In conclusion, given the complexity of the overall charge map and an inherent uncertainty in assigning charge transitions to a specific DQD, we do not attempt to evaluate absolute hole occupancy in each dot. However, as reported in the main text, we use magneto-spectroscopy to establish the parity of specific ICTs and focus our investigation within the gate voltage space enclosed by the shaded area of Fig.~\ref{sfig:largemap}.

\section{\label{sec:input-output}Alternative origin of non-monotonic behaviour}
\begin{figure}[b]
    \centering
    \includegraphics[scale=0.8]{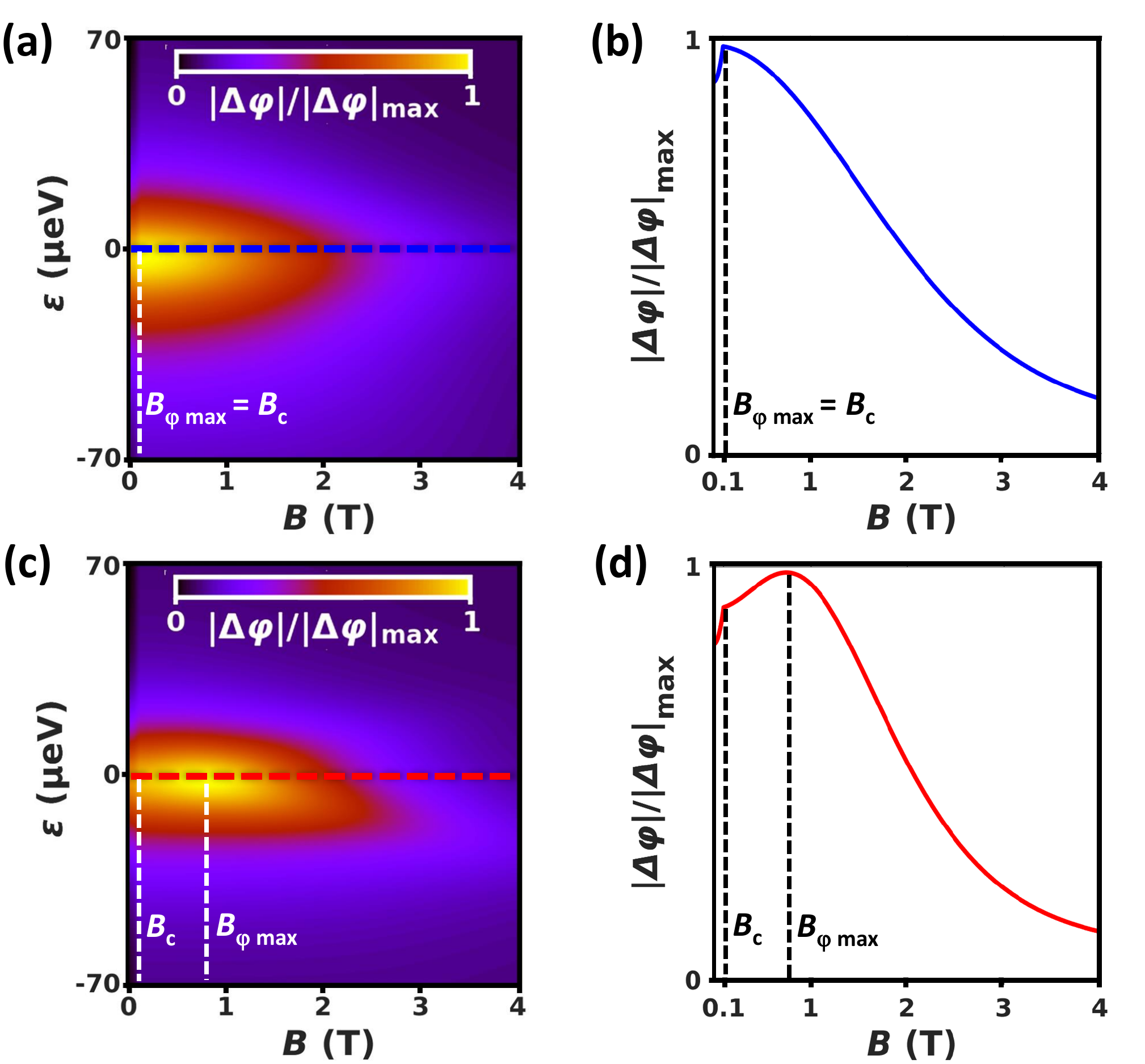}
    \caption{Results from the model based on input-output theory. (a) Phase response for equal g-factor values as a function of energy detuning and in-plane magnetic field for an even parity ICT calculated using Eq.~\ref{eq:Schroer}. Parameters used are: $f_\mathrm{o}= 341~\mathrm{MHz}$, $Q_\mathrm{c}=50$, $Q_\mathrm{i}=80$, $t= 10.6~\mathrm{\mu eV}$, $T_\mathrm{DQD} = 250~\mathrm{mK}$, $g_\mathrm{c}/(2\pi)=35~\mathrm{MHz}$, $g=g_\mathrm{L}+g_\mathrm{R}= 0.77$. The dashed white line highlights that $B_\mathrm{\varphi max} =B_\mathrm{c}=0.1~\mathrm{T}$. The horizontal dashed blue line highlights the zero detuning cut of panel (b). 
    (b) Phase response as a function of magnetic field for equal g-factors at $\varepsilon=0$. The dashed line highlights the maximum value of the phase response generated by the magnetic field dependence of $f_\mathrm{r}$, $Q_\mathrm{c}$ and $Q_\mathrm{i}$. (c) and (d) as per (a) and (b) respectively, but for the site-dependent g-factor values: $g_\mathrm{L}=0.49$ and $g_\mathrm{R}=0.28$.
    The vertical dashed lines highlight the local maxima due to a superconducting-to-normal transition ($B=B_\mathrm{c}$) and site-dependent g-factors ($B=B_\mathrm{\varphi max}$).}
    \label{sfig:in_out}
\end{figure}
In our experiments, there are other parameters that may be affected by a varying magnetic field ($B$), besides the investigated hole spin configurations in the DQD. Hence, it is worth directing some attention onto alternative origins of the $B-$field dependence. For example, part of the readout circuitry is made up of aluminum, namely the bond wires and bond pads used for interconnecting the silicon chip to a printed circuit board hosting the tank resonator. At the temperature of operation of the DQD, Al is superconductive in the absence of a $B-$field, but it is expected to turn to a normal metal state once a critical field threshold value is reached ($B_c$). We aim to demonstrate that the  non-monotonic dependence of the resonator response reported in the main manuscript does not trivially originate from a superconducting-to-normal transition in the circuitry interconnects. To this end, we use a model based on cavity input-output theory~\cite{schroer, petersson2012circuit, ezzo21}, as an alternative to the semi-classical approach based on quantum capacitance we used in the main text. With this approach, we are able to model the phase response as a function of the bare cavity resonant frequency $f_\mathrm{r}$, the coupling and internal quality factors $Q_\mathrm{c}$ and $Q_\mathrm{i}$, and the charge susceptibility $\chi$, a set of parameters that are affected by superconductive-to-normal transitions.\\\indent 
For our analysis we adapt the model presented in Ref.~\cite{schroer}, where the rf signal reflected by the device is given by: 
\begin{align}
    R = 1 + \frac{i\kappa}{\Delta_\mathrm{c}/\hbar - i(\kappa + \kappa_\mathrm{i})/2 + g_\mathrm{eff}\langle \chi \rangle}
    \label{eq:Schroer}
\end{align}
where $\Delta_\mathrm{c} = h(f_\mathrm{r}-f_\mathrm{o})$ is the detuning of the resonator from the measurement frequency $f_\mathrm{o}$, $\kappa=2\pi f_\mathrm{r}/Q_\mathrm{c}$ is the cavity decay rate caused by coupling to the cavity ports, $\kappa_\mathrm{i}= 2\pi f_\mathrm{r}/Q_\mathrm{i}$ is the decay rate due to internal loss mechanisms and $g_\mathrm{eff}=(2t/\Omega)g_\mathrm{c}$ is the effective resonator-charge coupling rate, where $\Omega = \sqrt{4t^2 +\varepsilon^2}$ and $g_\mathrm{c}$ is the bare coupling rate. The thermally averaged charge susceptibility $\langle \chi \rangle$ is proportional to the expression for $\Delta \varphi$ discussed in the main text. To model our situation we use the system parameters from the main text, with the addition of $g_c/(2\pi)=35~\mathrm{MHz}$ based on results reported from a similar hole-based system \cite{ezzo21}.  
We then consider a scenario where the aluminium undergoes a superconducting-to-normal transition at a critical field $B_\mathrm{c}= 0.1~\mathrm{T}$. This value can be considered an upper limit based on reports of critical field values for Al at mK temperatures~\cite{Caplan1965}. We include this transition in our model by making $f_\mathrm{r}$, $Q_\mathrm{c}$ and $Q_\mathrm{i}$ dependent on $B$ up to the critical field value. In brief, the superconducting-to-normal transition affects the resonator parameters so that for $B<B_\mathrm{c}$, $f_\mathrm{r}$
depends quadratically on $B$~\cite{petersson2012circuit, krollMag}, and for $B\geq B_\mathrm{c}$, $f_\mathrm{r}$ becomes constant and $Q_\mathrm{c}$ and $Q_\mathrm{i}$ experience a small step change.\\\indent
Figure \ref{sfig:in_out} compares the result of two variations of the model. In particular, panels (a) and (b) represent the case for which the Landé g-factors in the two dots are identical. One can observe a non-monotonic dependence of the phase signal caused by a superconducting-to-normal transition, which results in a peak at $B = B_\mathrm{c}$. In contrast, panels (c) and (d) show the case where the values of $g-$factor differ between dots. One can again observe non-monotonic behaviour but with distinct contributions arising from the superconductive-to-normal transition and the site-dependent $g-$factors. The latter effect results in a separate peak at $B = B_\mathrm{\varphi max}$, with the exact value depending on the details of the DQD physics, as reported in the main text.\\\indent
We conclude that the $B-$field dependence of the phase response due to a trivial superconducting-to-normal metal transition in the circuit interconnects can be easily distinguished from the effect due to site-dependent $g-$factors, as long as $B_\mathrm{\varphi max}$ and $B_\mathrm{c}$ have clearly distinct values. Although we may have not observed the trivial peak at $B=B_\mathrm{c}$ in our experiments due to lack of resolution in the step size at the few mT range, the observed large value for $B_\mathrm{\varphi max}$ indicates that its origin is incompatible with a superconducting-to-normal metal transition.

\end{document}